\documentclass{svproc}
\usepackage{url}
\usepackage{wrapfig}
\usepackage{lipsum} 
\usepackage{graphicx}
\usepackage[T1]{fontenc}
\usepackage[utf8]{inputenc}
\usepackage{lmodern}
\usepackage{caption}
\usepackage{wrapfig}
\usepackage{chngcntr}
\usepackage{pdflscape}
\usepackage[lastpage,user]{zref}
\usepackage{lipsum}

\begin{document}
\mainmatter              % start of a contribution
%
%\linenumbers
\title{CP violation in $D^0 \rightarrow K_S^0 K_S^0$ decay at Belle }
\titlerunning{CP violation in $D^0 \rightarrow K_S^0 K_S^0$ decays}  % abbreviated title (for running head)
\author{N. Dash\inst{} \\(On
behalf of the Belle Collaboration)}

\authorrunning{N. Dash}

\institute{Indian Institute of Technology Bhubaneswar, INDIA\\
\email{nd11@iitbbs.ac.in} }%,\\ WWW home page:}
%\texttt{http://users/\homedir iekeland/web/welcome.html}
%\and
%Universit\'{e} de Paris-Sud,
%Laboratoire d'Analyse Num\'{e}rique, B\^{a}timent 425,\\
%F-91405 Orsay Cedex, France}

%\institute{\small  Indian Institute of Technology Bhubaneswar, INDIA \\nd11@iitbbs.ac.in
%\and \small  Indian Institute of Science Education and Research Mohali, INDIA\and \small  High Energy Accelerator Research Organization (KEK) Tsukuba, JAPAN}
%}

\maketitle              % typeset the title of the contribution

\begin{abstract}
We report the preliminary measurement of the $CP$ asymmetry of the $D^0 \rightarrow K^0_S K^0_S$ decay, which is consistent  with no $CP$ violation and improves the uncertainty with respect to the previous measurement of this quantity by more than a factor of three. Also, the expected  precision on $CP$ asymmetry of the $D^0 \rightarrow K^0_S K^0_S$ decay by the Belle II experiment is presented.
% We would like to encourage you to list your keywords within
% the abstract section using the \keywords{...} command.
\keywords{Belle, $CP$ asymmetry, $D^{0}$ meson}
\end{abstract}

\section{Introduction}
Charge-conjugation Parity Violation (CPV) in charmed meson decays has not yet been observed and is predicted to be (10$^{-3}$) in the Standard Model (SM). The combined result of $\Delta A_{CP}$ $(A _{CP}^{ D^{0} \rightarrow \pi^{+} \pi^{-}}$ $-$  $A_{CP}^{D^{0} \rightarrow K^{+} K^{-}}$) by HFAG~\cite{99} gives an agreement with no CPV at 9.3\% confidence level (CL). Though there is no current evidence of non-zero asymmetry, CPV in charm decay is investigated in other channels. Singly Cabibbo-suppressed (SCS) decays are of special interest as the possibility of interference with NP amplitudes could lead to larger CPV than predicted by SM, the $D^0 \rightarrow K^0_S K^0_S$  decay is one such channel~\cite{hiller}. 
The $D^0$ candidates are selected as coming from the decay $D^{*+} \rightarrow D^0\pi^+_s$, where $\pi^+_s$ denotes the low-momentum "slow" pion whose charge  reveals the flavor content of neutral $D$ meson ($D^0$ or $\bar{D}^0$ ) at its production vertex. A stringent selection criterion is applied on the momentum of the $D^{*+}$ candidate in the $e^+e^-$ center-of-mass frame, $p^*(D^*)$, to suppress $D^{*+}$  coming from $B$ decays as well as to reduce the combinatorial background. The extracted raw asymmetry is given by:$ A_{\rm raw} = \frac{N(D^{0}\rightarrow f)- N(\bar D^{0}\rightarrow \bar f)}{N(D^{0}\rightarrow f)+ N(\bar D^{0}\rightarrow {\bar {f}})} =  A_{CP} + A_{FB} + A^{\pm{}}_{\epsilon}$. Here, $A_{FB}$ is the forward-backward production asymmetry, and $A^{\pm{}}_{\epsilon}$ is the asymmetry due to different detection efficiencies for positively and negatively charged pions. The $A_{CP}$ of signal mode (S) is measured relative to other well measured normalization decay mode (N). Such  an  approach  enables  the  cancellation  of  several  sources  of  systematic  uncertainties that are common to both the signal and normalisation modes. The $CP$ asymmetry of the signal mode can then be expressed as: $A_{CP}(S) = A_{\rm raw}(S) - A_{\rm raw}(N) + A_{CP}(N)$. For $A_{CP}(N)$, the world average value \cite{pdg} is used. The result of the analysis presented here is based on  data sample corresponding to 921 fb$^{-1}$ integrated luminosity collected with the Belle detector at the KEKB asymmetric energy $e^{+}e^{-}$ collider at center of mass energy $\sqrt{s}$ $\approx$ 10.58 GeV. ~\cite{abashian}. The most recent SM-based analysis obtained a 95\% CL upper limit of 1.1\% for direct $CPV$ in $D^{0} \to K^0_{S} K^0_{S}$ decay~\cite{Nierste:2015zra}. The search for $CP$ asymmetry in $D^{0} \to K^0_{S} K^0_{S}$ has been performed first by the CLEO ~\cite{Bonvicini:2000qm} as $(-23 \pm 19)\%$ and recently by LHCb as ($-$2.9 $\pm$ 5.2 $\pm$ 2.2)\%, where the first uncertainty is statistical and the second is systematic~\cite{Aaij:2015fua} and result is consistent with no $CPV$, in agreement with SM expectations. 
\begin{wrapfigure}{R}{0.47\textwidth}
\centering
\includegraphics[width=6cm,height=3.0cm]{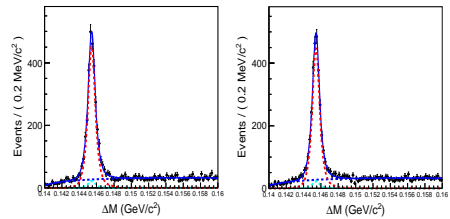}
\caption{The $\Delta M$ distribution for the $D^0 \rightarrow K_S^0 K_S^0$.}
\label{fig:markern}
\end{wrapfigure}
\section{Signal extraction and Systematic uncertainties}
 A simultaneous fit of the $\Delta M$ (the mass difference between reconstructed $D^*$ and $D^0$) for $D^{*+}$ and $D^{*-}$ is used (Figure~\ref{fig:markern}) to estimate the asymmetry. Here the normalization mode is $D^0 \to K_S^0 \pi^0$.  The signal yield for $D^0 \to K_S^0 K_S^0$ is $5399 \pm 87$ events and for $D^0 \to K_S^0 \pi^0$ is $531807 \pm 796$ events. The $A_{\rm raw}$ observed in data for signal and normalization modes are $(+0.45 \pm 1.53)\%$ and $(+0.16 \pm 0.14)\%$, respectively. The total systematic uncertainty for the  $A_{CP}$ in the $D^{0}\rightarrow K_{S}^{0}K_{S}^{0}$ decay is $\pm$0.17\%~\cite{nibedita}. 
%.................................................................
\section{Result and Conclusion}
The measured time-integrated $CP$-violating asymmetry in the $D^{0}\rightarrow K_{S}^{0}K_{S}^{0}$ decay is found to be $A_{CP}$ = ($-$0.02 $\pm$ 1.53 $\pm$ 0.17)\%. The dominant systematic uncertainty comes from the $A_{CP}$ error of the normalisation mode which is consistent with SM expectations and is a significant improvement compared to the previous results from CLEO~\cite{Bonvicini:2000qm} and LHCb ~\cite{Aaij:2015fua}, already probing the region of interest. The Belle II experiment is designed to record data at SuperKEKB, a major upgrade of KEKB, expect $A_{CP}$ with a precision of 0.2\% and should be able to test the SM. 
%\begin{figure}
%\centering

%\end{figure}

\end{document}